\documentclass[]{spie}  %>>> use for US letter paper
\usepackage{amsmath,amsfonts,amssymb}
\usepackage{graphicx}
\usepackage[colorlinks,allcolors=black,pdftex]{hyperref}
\usepackage{chemformula}
\usepackage{siunitx}
\usepackage{textcomp}
\usepackage{adjustbox}

\title{CCAT-prime: The Design and Characterization of the Silicon Mirrors for the Fabry-Perot Interferometer in the Epoch of Reionization Spectrometer}

\author[a]{Bugao Zou}
\author[b,c]{Steve K. Choi}
\author[d]{Nicholas F. Cothard}
\author[c]{Rodrigo Freundt}
\author[b]{Zachary B. Huber}
\author[b,f]{Yaqiong Li}
\author[b]{Michael D. Niemack}
\author[e]{Thomas Nikola}
\author[g]{Dominik A. Riechers}
\author[e]{Kayla M. Rossi}
\author[c]{Gordon J. Stacey}
\author[b]{Eve M. Vavagiakis}
\author[ ]{the CCAT-prime collaboration}

\affil[a]{Department of Applied and Engineering Physics, Cornell University, Ithaca, NY 14853, USA}
\affil[b]{Department of Physics, Cornell University, Ithaca, NY 14853, USA}
\affil[c]{Department of Astronomy, Cornell University, Ithaca, NY 14853, USA}
\affil[d]{NASA Goddard Space Flight Center, Greenbelt, MD 20771, USA}
\affil[e]{Cornell Center for Astrophysics and Planetary Sciences, Ithaca, NY 14853, USA}
\affil[f]{Kavli Institute at Cornell for Nanoscale Science, Cornell University, Ithaca, NY 14853, USA}
\affil[g]{I. Physikalisches Institut, Universit\"at zu K\"oln, Z\"ulpicher Strasse 77, D-50937 K\"oln, Germany}

\authorinfo{Further author information: (Send correspondence to B.Z.)\\B.Z.: E-mail: bz332@cornell.edu, Telephone: 1 571 299 0238}

\begin{document} 
\maketitle

\begin{abstract}
The Epoch of Reionization Spectrometer (EoR-Spec) is one of the instrument modules to be installed in the Prime-Cam receiver of the Fred Young Submillimeter Telescope (FYST). This six-meter aperture telescope will be built on Cerro Chajnantor in the Atacama Desert in Chile. EoR-Spec is designed to probe early star-forming regions by measuring the [CII] fine-structure lines between redshift z = 3.5 and z = 8 using the line intensity mapping technique. The module is equipped with a scanning Fabry-Perot interferometer (FPI) to achieve the spectral resolving power of about $RP$ = 100. The FPI consists of two parallel and identical, highly reflective mirrors with a clear aperture of 14 cm, forming a resonating cavity called etalon. The mirrors are silicon-based and patterned with double-layer metamaterial anti-reflection coatings (ARC) on one side and metal mesh reflectors on the other. The double-layer ARCs ensure a low reflectance at one substrate surface and help tailor the reflectance profile over the FPI bandwidth. Here we present the design, fabrication processes, test setup, and characterization of silicon mirrors for the FPI.
\end{abstract}


\section{INTRODUCTION}
\label{sec:intro}  % \label{} allows reference to this section
The CCAT-prime Collaboration is constructing the Fred Young Submillimeter Telescope (FYST) on Cerro Chajnantor in the Atacama Desert in Chile, a wide-field, 6-m aperture telescope optimized to operate in the millimeter to submillimeter (100 GHz to 1.5 THz) range.\cite{parshley2018ccat,aravena2019ccat} The Epoch of Reionization Spectrometer (EoR-Spec) is a spectrometer instrument module for the Prime-Cam receiver of the telescope.\cite{vavagiakis2018prime,cothard2020design,ThomasSPIE} It is designed to measure the 158 \unit{\um} [CII] fine-structure lines between redshift z = 3.5 and z = 8, or 210 to 420 GHz in frequency space, using line intensity mapping. A cryogenic, scanning Fabry-Perot interferometer (FPI) with silicon-substrate-based (SSB) mirrors will be installed in EoR-Spec to act as a narrow bandpass filter. The spectral resolving power requirement is about $RP=100$ across the band. The detector array will split the frequency coverage into two bands, one from 210 to 315 GHz and the other from 315 to 420 GHz.\cite{cothard2020design,ThomasSPIE} EoR-Spec will observe two orders of the FPI simultaneously to increase the mapping speed.%, which sets a tight constraint on the finesse distribution of the FPI.

FPIs are commonly used for astronomical observations in submillimeter\cite{bradford2002spifi,oberst2009submillimeter} to far-IR\cite{poglitsch1991mpe,douthit2018development,1991ApJ,latvakoski1999kuiper} bands for their high throughput and excellent spatial multiplexing.\cite{vaughan2017fabry,swain1998design} This type of interferometer utilizes multiple reflections between two parallel reflecting surfaces to form a resonating cavity called etalon. A traditional FPI used in the far-IR to millimeter regime adopts free-standing metal meshes as reflectors, whose spectral response is determined by the geometry of the meshes.\cite{sakai1983far} There are two types of standard metal meshes. One is the inductive mesh (Figure \ref{fig:sch}c) acting as a high pass filter, usually formed from an electroforming process. The other one is the complementary capacitive mesh, which behaves like a low pass filter and is usually fabricated by vacuum evaporating metals on a thin dielectric substrate. The finesse of free-standing meshes is strongly related to frequency as $F \propto \nu^{-2.5 \ \text{to} \ 3}$,\cite{vavagiakis2018prime} thus the bandwidth where the FPI can suitably work within is limited. Free-standing metal meshes with a large diameter, if used for EoR-Spec with a 14 cm clear aperture, are also susceptible to vibrations because they are usually very thin ($\sim \lambda/3$) and have no or thin substrates.\cite{Cothard2018OptimizingTE}

The SSB FPI we designed aggregates current technology of both metal meshes and anti-reflection coatings (ARC). Each FPI mirror is made of a silicon wafer with metal meshes patterned on one side and double-layer ARCs etched on the other. The ARC structure is needed to mitigate the Fresnel reflections of silicon. The regular method of creating ARCs by coating the target surface with layers of dielectric films with different refractive indices\cite{rosen2013epoxy,lau2006millimeter} cannot be easily applied for cryogenic usage due to the mismatch of coefficients of thermal expansion (CTE) and limited bandwidth. We adopt the method of forming double-layer subwavelength metamaterial structures on one side of silicon. It solves the thermal expansion issue and provides a broader bandwidth than the single-layer ARCs.\cite{Cothard2018OptimizingTE} There are many fabrication processes proposed to create multi-layer ARCs, including the dicing saw method\cite{datta2013large,nitta2017broadband,young2017broadband}, laser machining method\cite{young2017broadband,d2001laser,matsumura2016millimeter}, and dry etching method\cite{Cothard2018OptimizingTE,gallardo2017deep,hasebe2021fabrication,defrance20181}. We chose the dry etching method, which uses a bombardment of plasma ions to construct target structures with a high aspect ratio of up to 50:1 and allows for highly controllable processing at a submicron scale.

The reflectance and thus finesse profile of the mirror is determined by the geometry of both the mesh and ARC structure. The mature semiconductor technology enables us to precisely tune the dimensions for both structures, thus allowing for fine adjustments for the resolving power profile of the FPI across the band. Using silicon wafers as substrates also dramatically enhances the mechanical stability of the instrument. Additionally, the low loss tangent and low CTE of a high purity silicon wafer ensure high throughput even in cryogenic conditions.

The paper is outlined as follows. In Sect.~\ref{sec:design}, we discuss the design of the FPI mirror, and in Sect.~\ref{sec:fab}, we present the fabrication processes for it. Later in Sect.~\ref{sec:char} we talk about how we characterized our sample wafers and present current measurements for them. Finally, in Sect.~\ref{sec:future} we conclude with the plans for the next steps and additional work in progress with similar SSB techniques.

\section{Design} \label{sec:design}
\subsection{General Requirements for the Design} \label{sec:design_1}

\begin{figure}
\centering
\includegraphics[width=\textwidth]{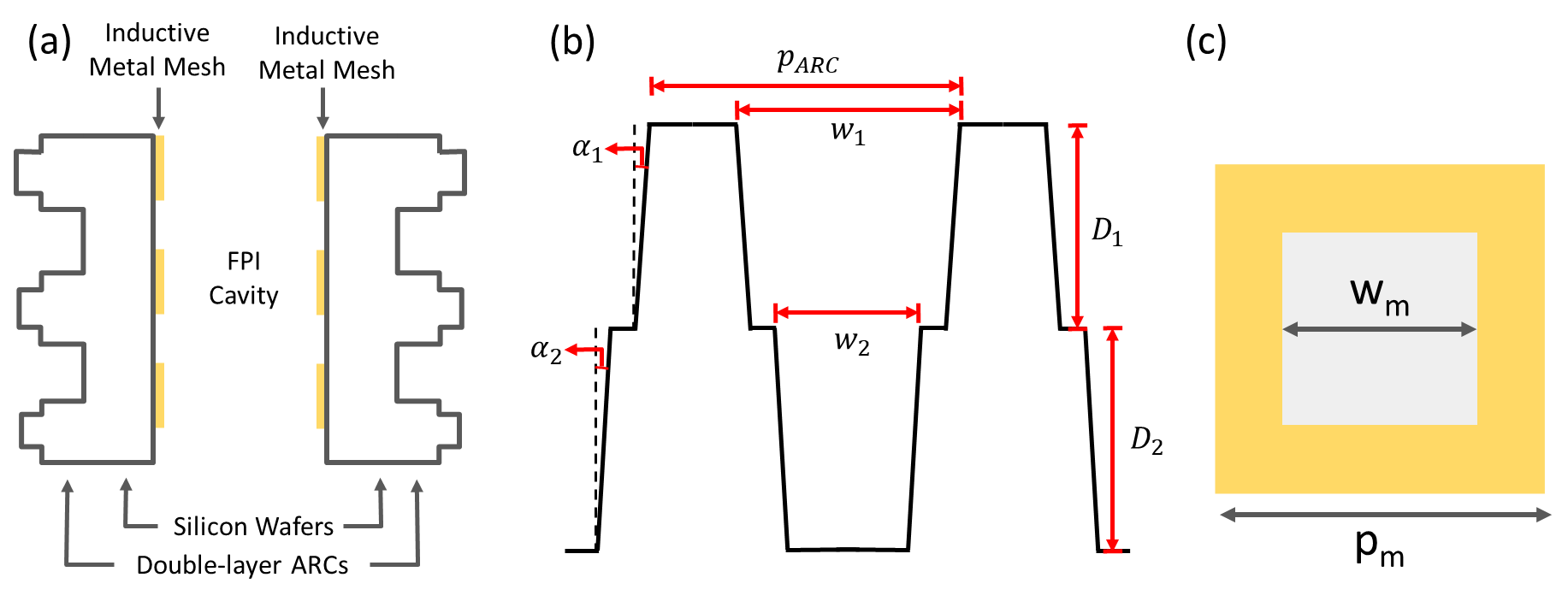}
\caption{Schematics of (a) the EoR-Spec FPI\cite{cothard2020design}, (b) the double-layer ARC structure showing the parameter acronyms used in the text and (c) the unit cell of the inductive metal mesh structure where the yellow color indicates the metal portion.}
\label{fig:sch}
\end{figure}

The structure of the EoR-Spec FPI is shown in Figure \ref{fig:sch}a. The double-layer ARC structure and the metal meshes are patterned on either side of the FPI mirror. As mentioned in Sect.~\ref{sec:intro}, EoR-Spec will map two spectral orders of the FPI simultaneously. More specifically, the FPI will be used in second order in the 210 to 315 GHz band and third order in the 315 to 420 GHz band. To match the spectral resolution requirement of about $RP$ = 100, we aim for finesse $35 < F < 60$ for the second-order band and finesse $27 < F < 40$ for the third-order band, corresponding to $70 < RP < 120$ and $81 < RP < 120$ respectively with $RP = kF$ where $k$ is the spectral order. The finesse of the FPI is further related to the reflectance of FPI mirrors $R$ by $F=\pi \sqrt{R}/{(1-R)}$.

%The double-layer ARC structure mitigates the Fresnel reflections of silicon and provides a broader bandwidth than the single-layer ARC.\cite{Cothard2018OptimizingTE} The reflectance profile of the mirror is determined by the geometry of both the mesh structure and the ARC structure, as is shown in Figure \ref{fig:sim1}b. The topic will be covered in the following subsections.

If the absorptance from the silicon substrate and metal meshes is neglected, the peak transmittance of an FPI fringe is determined by the formula $(1-r_1)(1-r_2)/(1-\sqrt{r_1 r_2})$, where $r_1$ and $r_2$ are the reflectances of the two FPI mirrors respectively.\cite{ismail2016fabry} It suggests that power leakage will be introduced if the reflectance profiles of the two mirrors constructing the FPI do not match, which is confirmed by our electromagnetic simulations of the FPI performance. Therefore, we use the same ARC and metal mesh design for both mirrors to minimize leakage. % It may seem that the transmittance profile of the ARC does not matter as long as we keep the dimensions the same for both mirrors, but 

\begin{figure}
\centering
\includegraphics[width=\textwidth]{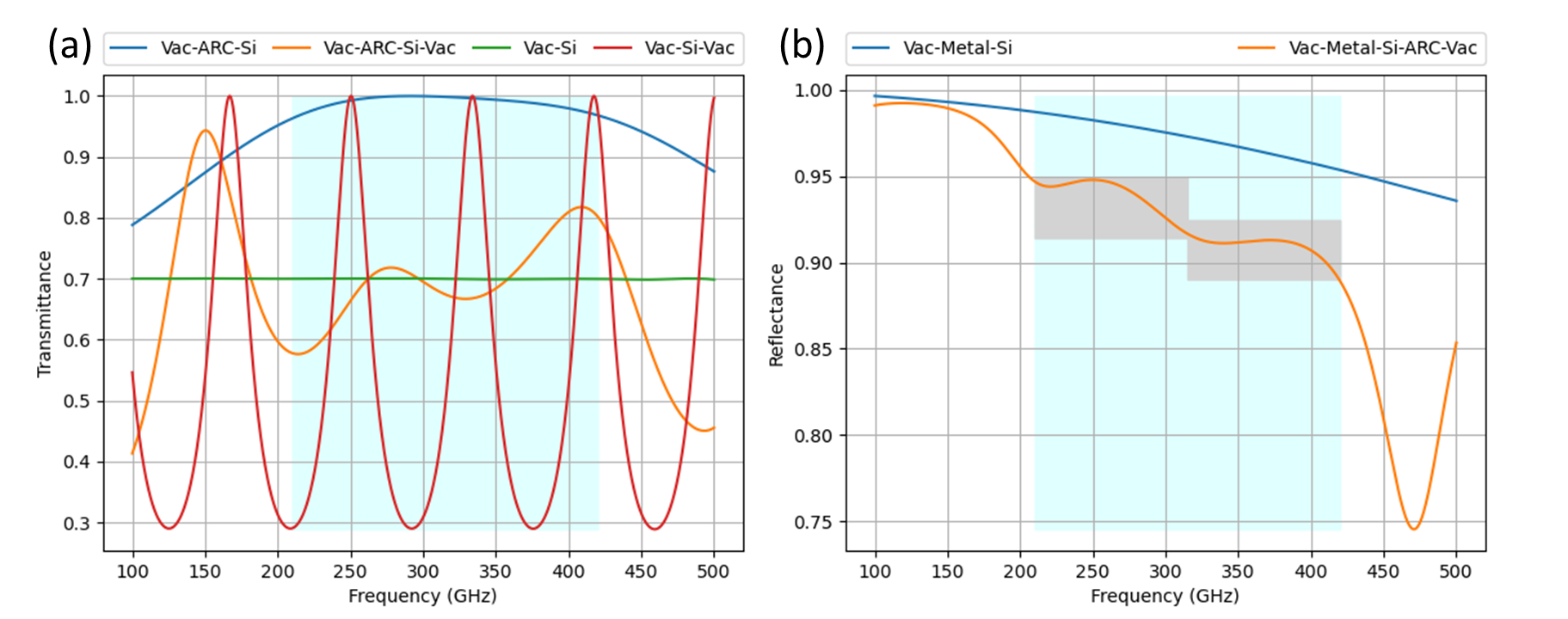}
\caption{CST-MWS calculations of (a) the transmittance and (b) the reflectance profiles of different interfaces for the optimized ARC and metal mesh design. The thickness of silicon wafers is set to be $t=\SI{525}{\um}$ for all simulations, which is the nominal thickness for the silicon substrates discussed in the paper. The cyan boxes in (a) and (b) indicate the frequency range of interest for us (210 to 420 GHz). The grey-shaded boxes in (b) define the reflectance limit in the second- and third-order band respectively, as per the finesse requirement.}
\label{fig:sim1}
\end{figure}

Figure \ref{fig:sim1}a shows the transmittance profiles of essential interfaces for the ARC design. The vacuum-silicon interface (Vac-Si) indicated by the green curve has a uniform transmittance of around 70\% across the band of interest. A bare silicon wafer presents strong parasitic resonances whose transmittance oscillates from about 30 \% to 100 \%, as suggested by the red curve. The period of the resonances is determined by the thickness of the wafer. One side of the silicon wafer is patterned with the ARC structure to reduce the resonances. The blue curve shows the transmittance of the vacuum-ARC-silicon interface (Vac-ARC-Si) for the optimized design, which stays above 96 \% for the whole band. However, since the other side is still bare silicon and the ARCs are not perfect, the transmittance of the entire ARC coated wafer shown by the orange curve still oscillates around the green curve, i.e., the Vac-Si transmittance profile. The amplitude of the parasitic resonances is negatively related to the effectiveness of the ARC structure (transmittance of the Vac-ARC-Si interface). The wavy features may seem unwanted at first glance, but it provides a powerful way to tailor the reflectance profile of the FPI mirror. The typical reflectance profile of an inductive metal mesh is shown by the blue curve in Figure \ref{fig:sim1}b. It is difficult to tune the frequency dependency of the profile by changing the dimensions of the metal meshes. However, it can be tailored to fit the grey-shaded boxes, which define the resolution requirement of the instrument by the equation mentioned above, by properly tuning the dimensions for the ARC structure. That being said, we would not optimize for maximum transmittance for the ARC structure, but need to tune different parameters to find optimal resonating shapes to match the requirement for the reflectance as indicated by the grey-shaded boxes.

We also note that the whole design is optimized at 4 K, the cryogenic temperature at which EoR-Spec will operate. Specifically, we adopt the dielectric constant of high resistivity silicon $\varepsilon_{\ch{Si}} = 11.5$\cite{datta2013large}, the resistivity of gold $\rho_{\ch{Au}} = \SI{2.2d-10}{\ohm \metre}$ and resistivity of chromium $\rho_{\ch{Cr}} = \SI{1.6d-8}{\ohm \metre}$.\cite{lide2004crc} These two metal materials are used for the fabrication of metal mesh structure.

\subsection{Initial Parameters of the ARC Design}
The structure of the double-layer ARCs is shown in Figure \ref{fig:sch}b. It can be characterized by the pitch size $p_\text{ARC}$, the depth of the upper and lower layer $D_1$ and $D_2$, the width of the hole structures in these two layers $w_1$ and $w_2$, and the corresponding taper angles $\alpha _1$ and $\alpha _2$. The tapering sidewalls are caused by plasma etchings during the fabrication processes. The ARC structure can be treated as a stack of three homogeneous layers with different refractive indices $n_1$, $n_2$ and $n_{\ch{Si}}$. If the thickness of the two ARC layers is set to be $\lambda/4$ to form destructive interferences, where $\lambda$ is the wavelength inside the layer material, the reflectance at normal incidence of the triple-layer structure is given by
\begin{equation}
R=\left(\frac{n_{0} n_{2}^{2}-n_{\ch{Si}}n_{1}^{2}}{n_{0}n_{2}^{2}+n_{\ch{Si}}n_{1}^{2}}\right)^{2}
\end{equation}
from the theory of layered dielectrics, with $n_0$ the refractive index of the medium (in our case, vacuum).\cite{yeh1990optical,cothard2021cryogenic} To minimize the reflection, $\left(n_{2}/n_{1}\right)^{2}=\varepsilon_{2}/\varepsilon_{1}=n_{\ch{Si}}/n_{0}=3.4$ is attained where $\varepsilon_{1}$ and $\varepsilon_{2}$ are the dielectric constants for the upper and lower ARC layer. However, the relation only minimizes the reflection at the center of the band since the thickness of each ARC layer is optimized for the central frequency. To maximize the bandwidth of the ARC structure with high transmittance, we use Computer Simulation Technology Microwave Studio (CST-MWS) to simulate the transmittance profile of the triple layer stack (Vac-ARC-Si) and sweep over the parameter space for $\varepsilon_{1}$ and $\varepsilon_{2}$. We later chose $\varepsilon_{1}=2.0$ and $\varepsilon_{2}=5.7$ as the initial design for its broadband high transmittance coverage. The thickness of these two layers is $D_1 = \SI{168.2}{\um}$ and $D_2 = \SI{99.7}{\um}$ using the $\lambda/4$ relation, noting that the central frequency 315 GHz is used for wavelength calculation.

According to the equivalent circuit model theory, the dielectric constant of the ${i}$-th ARC layer is related to its geometry (neglecting the tapering sidewalls) by 
\begin{equation}
\varepsilon_{\text {ARC}}^{i}=\varepsilon_{\ch{Si}}\left[1-k_{i}+\left(\frac{1-k_{i}}{k_{i}}+\frac{\varepsilon_{\ch{Si}}}{\varepsilon_{0}}\right)^{-1}\right]
\end{equation}
where $k_{i}=w_{i}/p_{\text {ARC}}$ and $\varepsilon_{0}$ is the dielectric constant for vacuum.\cite{cothard2021cryogenic} We get $k_1 =0.913$ and $k_2 =0.588$ after solving for the equation. The pitch of the ARC should be as large as possible to make fabrication easier. However, the pitch size needs to be significantly less than the wavelengths at which the FPI operates, since when it becomes comparable to the wavelengths severe diffraction effects occur. After a few test runs of CST-MWS simulations, we set $p_{\text {ARC}}$ to be \SI{140}{\um}, thus $w_1 = \SI{127.8}{\um}$ and $w_2 = \SI{82.3}{\um}$ correspondingly.

For now, we have not considered the physical taper angles caused by the fabrication processes, which are found to be about $\alpha_1 = \ang{0.5}$ and $\alpha_2 = \ang{1.5}$ according to our pathfinder wafers. All the dimensions above form the initial parameter set for the double-layer ARC model. Further optimization is needed after having the initial metal mesh design, which together with the ARC design aims to present the reflectance profile we want.
% As we mentioned above, we want to optimize the shape of the transmittance/reflectance profile of the ARC structure to tailor the reflectance profile of the whole FPI mirror. For now, we have not considered the taper angles in the ARC structure. Nonetheless, we have to redo the CST-MWS simulation for the whole FPI mirror later once the mesh dimension is roughly determined. The tapered sidewalls in ARC layers will be taken into account then.

\subsection{Initial Parameters of the Metal Mesh Design}
As mentioned in Sect.~\ref{sec:intro}, two types of metal meshes are widely used in the field: the inductive mesh, shown in Figure \ref{fig:sch}c having a pitch $p_m$ and an opening width $w_m$, which possesses high pass filtering properties and the complementary capacitive mesh behaving like a low pass filter.
%There are two types of metal mesh filters commonly used in the field. One is the inductive mesh displayed in Figure \ref{fig:sch}c having a pitch $p_m$ and an opening width $w_m$. The other one is the complementary capacitive mesh, which interchanges the position of the metal and the blank space. The former mesh acts as a high pass filter, while the latter behaves like a low pass filter. 
Since the EoR-Spec FPI will work in two different orders between 210 and 420 GHz with roughly the same resolving power, the finesse range and thus the reflectance range should decrease as frequency increases, as shown by the grey-shaded boxes in Figure \ref{fig:sim1}b. Hence, we choose to use the inductive mesh design to confine the reflectance profile to be decreasing. Gold is selected as the mesh material for its low resistivity, thus low ohmic losses and superb chemical stability. The metal mesh is modeled as a stack of a 10 nm chromium layer needed for gold adhesion on the silicon substrate and a 100 nm gold layer in CST-MWS. The shape of the reflectance profile can be coarsely tuned by changing $p_m$ and $w_m$. The larger the opening width, the lower the reflectance. The pitch parameter behaves the opposite way. For the convenience of arranging both the ARC and mesh structure in the same unit cell for further simulations, $p_\text{ARC}$ can be set to be integer multiples of $p_m$. For the initial mesh design, we use $p_m = \SI{140}{\um}$ and $w_m = \SI{70}{\um}$, whose reflectance profile is above the target grey-shaded region in Figure \ref{fig:sim1}b.

\begin{figure}
\centering
\includegraphics[width=\textwidth]{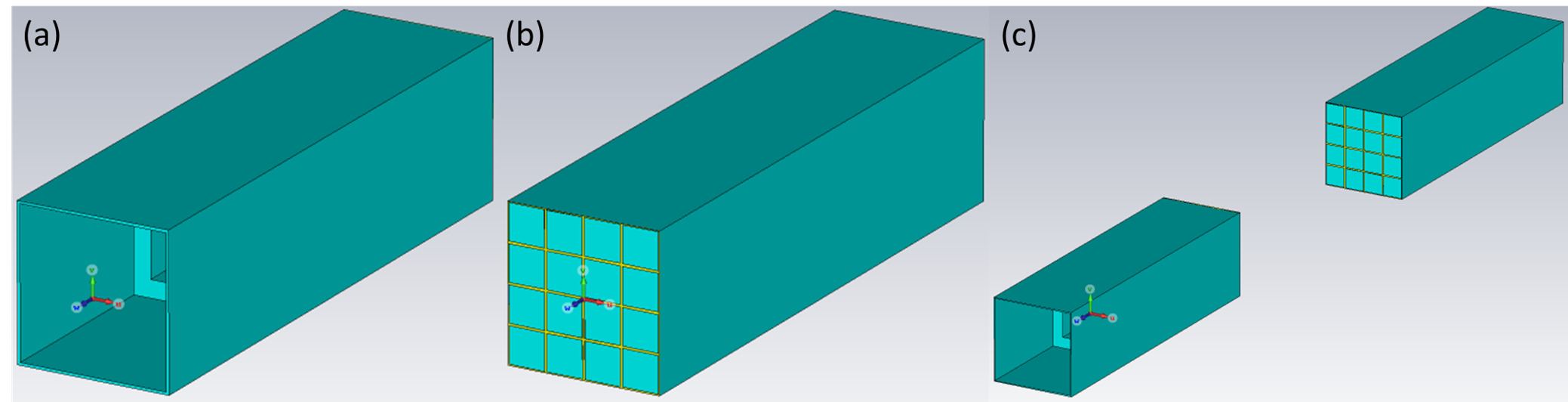}
\caption{The unit cell used in CST-MWS simulations. (a) and (b) are two isometric views for the unit cell of the single FPI mirror, showing the ARC and mesh structure on either side. (c) is the unit cell for the full FPI containing two identical FPI mirrors with a certain mirror separation.}
\label{fig:unit}
\end{figure}

\subsection{Optimize for the FPI mirror and the Full FPI}
Once having the initial parameters for both structures, we can construct the unit cell of the FPI mirror for CST-MWS simulations, shown in Figure \ref{fig:unit}a and \ref{fig:unit}b. Then we sweep over all these parameters except for $p_\text{ARC}$, $\alpha_1$ and $\alpha_2$ (also limit $p_m$ to be divisors of $p_\text{ARC}$) to fit the overall reflectance curve to the resolution requirement. The figure of merit we optimized is the bandwidth of the reflectance profile enclosed by the grey-shaded resolution targeted region, called the reflectance bandwidth. Table \ref{tab:Data} lists the optimized set of parameters we used for the practical fabrication processes, with a maximized reflectance bandwidth of 210 GHz. The corresponding reflectance profile is shown by the orange curve in Figure \ref{fig:sim1}b. The tolerance for each parameter can be calculated as the range where the reflectance bandwidth is greater than 190 GHz. The tolerance set is also shown in Table \ref{tab:Data}.

% \begin{tiny}

\begin{table}[ht]
\caption{The design parameters and corresponding tolerances of the FPI mirror and lab measurements for selected samples with standard deviations. Most parameter acronyms are defined in Figure \ref{fig:sch}(b) and (c) while some in the text.}
\label{tab:Data}
\begin{center} 
\begin{adjustbox}{max width=\textwidth}
\begin{tabular}{cccccccc}
\hline
 & $D_1$ (\unit{\um}) & $D_2$ (\unit{\um}) & $w_1$ (\unit{\um}) & $w_2$ (\unit{\um}) & $w_m$ (\unit{\um}) & $\alpha_1$ (°) & $\alpha_2$ (°)\\
\hline
Optimized Design & $177.6$ & $107.3$ & $135.2$ & $99.4$ & $32.9$ & $0.5$ & $1.5$ \\
Tolerances & $(-4.0,+24.0)$ & $(-9.5,+8.8)$ & $(-2.4,+1.3)$ & $(-2.0,+7.5)$ & $(-0.1,+0.3)$ & $(-0.5,+0.55)$ & $(-1.5,+1.35)$\\
% \hline
ARC Mirror & $178.8 \pm 1.0$ & $102.2 \pm 1.1$ & $137.5 \pm 0.6$ & $102.1 \pm 0.7$ & N/A & $0.10 \pm 0.06$ & $1.70 \pm 0.30$ \\
% \hline
FPI Mirror 1 & $180.5 \pm 0.7$ & $108.6 \pm 0.9$ & $135.0 \pm 0.4$ & $101.1 \pm 0.5$ & $32.85 \pm 0.03$ & $0.18 \pm 0.04$ & $0.98 \pm 0.23$ \\
% \hline
FPI Mirror 2 & $176.3 \pm 1.1$ & $107.7 \pm 0.8$ & $135.3 \pm 0.3$ & $100.8 \pm 0.6$ & $32.88 \pm 0.02$ & $0.15 \pm 0.07$ & $1.32 \pm 0.25$ \\
\hline
Fixed Parameters & \multicolumn{7}{c}{$p_\text{ARC}=\SI{140}{\um}$, $p_m =\SI{35}{\um}$, t = \SI{525}{\um}, $\varepsilon_{\ch{Si}}=11.5$}\\
\hline
\end{tabular}
\end{adjustbox}
\end{center}
\end{table}
% \end{tiny}

\begin{figure}
\centering
\includegraphics[width=\textwidth]{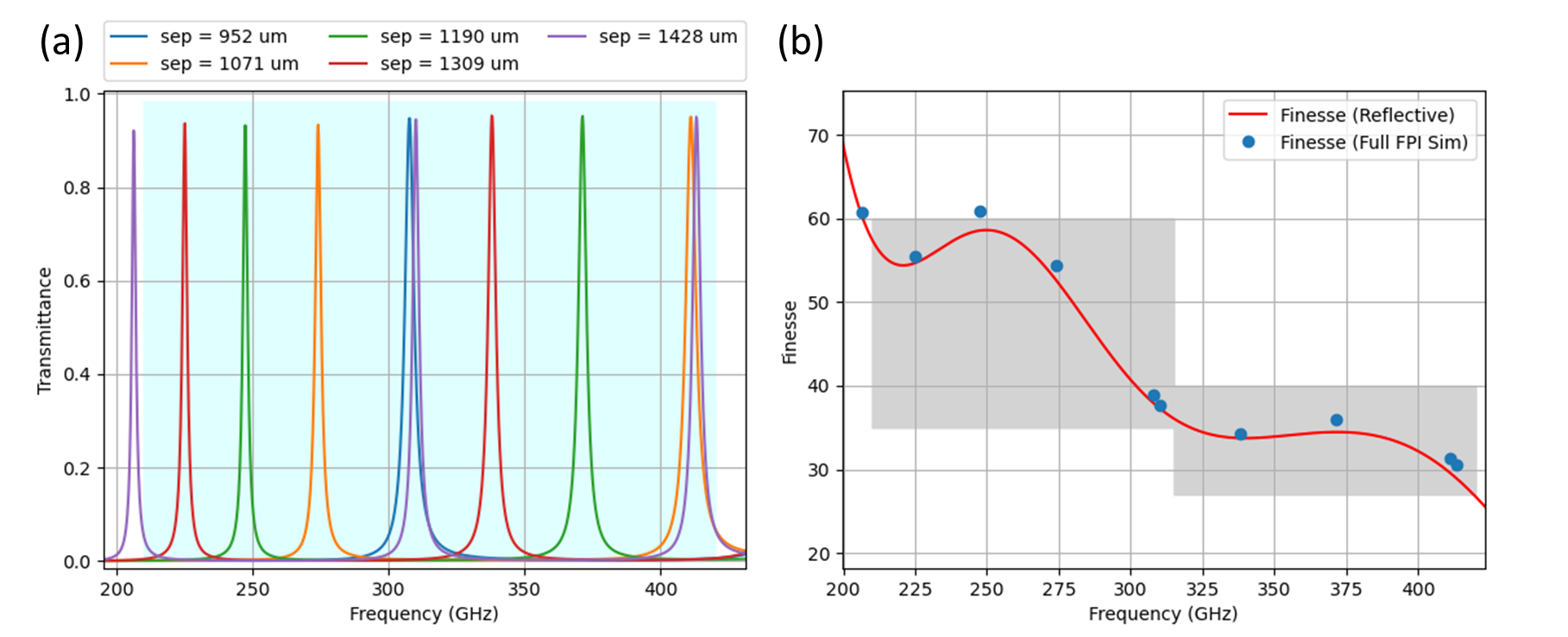}
\caption{(a) CST-MWS calculations of the transmittance fringes of the full EoR-Spec FPI at different mirror separations. (b) The finesse profile (red curve) derived from the reflectance profile and finesse data points (blue points) calculated from the full FPI CST-MWS simulations in (a). The grey-shaded boxes define the resolution requirement of the EoR-Spec instrument.}
\label{fig:sim2}
\end{figure}

To check the performance of the full FPI, we construct a new unit cell by placing two identical and optimized FPI mirrors separated by a certain mirror spacing, as shown in Figure \ref{fig:unit}c. By sweeping the mirror separation roughly over the operating range, from 952 to \SI{1309}{\um}, we can get the resonating fringes peaking at different frequencies (Figure \ref{fig:sim2}a). Multiple fringes corresponding to different spectral orders can be observed at a single mirror separation. The free spectral range $\delta \nu$ is determined by the separation $d$ via $\delta \nu = c/2d$, where $c$ is the speed of light. The peak transmittance of those fringes is not unity since the dielectric loss within the substrate (loss tangent of silicon $\tan \delta = \num{7d-5}$)\cite{datta2013large} and ohmic loss of the metal meshes are included in the simulation. The finesse of each resonating peak can be calculated by the definition $F = \nu / \delta \nu$. We then compared the calculated finesse data points with the finesse profile derived from the FPI mirror reflectance profile in Figure \ref{fig:sim2}b. They agree very well, verifying the simulation work for a single FPI mirror.

\section{Fabrication}
\label{sec:fab}

% \begin{itemize}
% \item ARC recipe
% \item Metal Mesh recipe
% \end{itemize}
\subsection{Substrate}
Double-sides polished $\langle100\rangle$ silicon wafers with 100 mm diameter and \SI{525}{\um} thickness were used for fabricating test samples. They are made from high-resistivity float-zone silicon specified to have resistivity $>$ \SI{10}{\kilo\ohm\cm}, orientation variation $<$ \ang{0.5}, thickness variation $<$ \SI{10}{\um} and bow/warp $<$ \SI{30}{\um}. High resistivity is desired to minimize dielectric loss within the substrate. Though the clear aperture for the EoR-Spec FPI is 14 cm, we chose to process 100 mm diameter wafers to demonstrate the fabrication technology and validate the design. We will adopt 200 mm wafers for the final FPI.

\subsection{Anti-reflection Coatings}
\begin{figure}
\centering
\includegraphics[width=\textwidth]{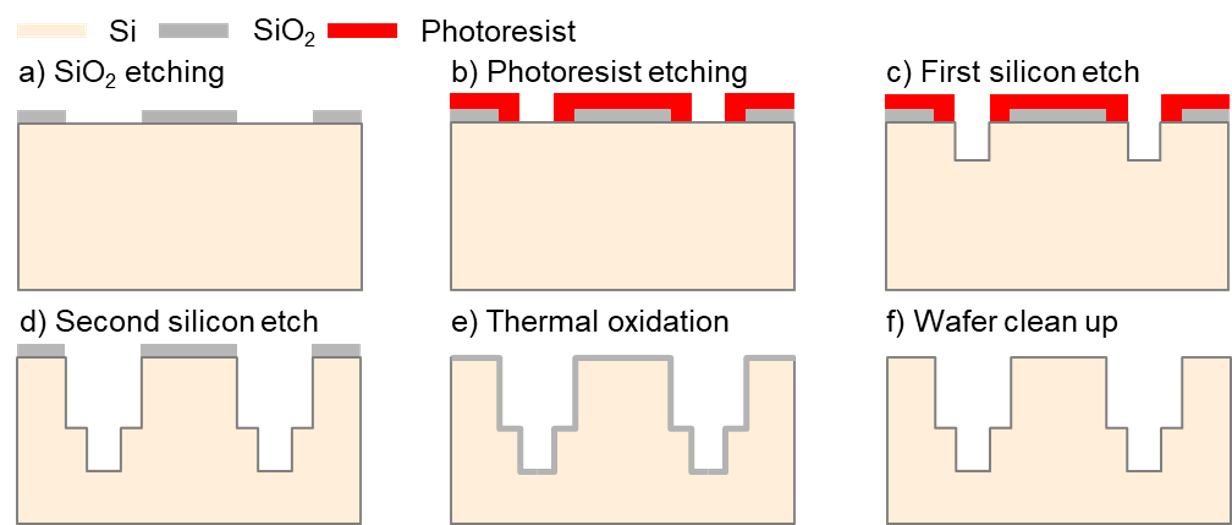}
\caption{Key fabrication processes of the double-layer ARC structure}
\label{fig:arc}
\end{figure}

The deep reactive-ion etching (DRIE) technique was employed to etch bulk silicon. It involves the Bosch processes which use alternate \ch{SF_6} and \ch{C_4 F_8} gas exposures to produce near-vertical sidewalls and high aspect ratio features.\cite{Chattopadhyay2017MicromachinedPF} The fabrication recipe of the double-layer ARC structure is outlined in Figure \ref{fig:arc}.

Before performing DRIE silicon etches, we patterned \ch{SiO_2} and photoresist grids as two silicon etch masks, defining the upper and lower geometry. 
%The thickness of those layers was determined by the selectivity of these materials to silicon.
Process (a) comprises three individual steps. The \ch{SiO_2} layer was first deposited onto the wafer by an Oxford plasma-enhanced chemical vapor deposition (PECVD) machine. Another layer of UV photoresist was then printed above the \ch{SiO_2} using an ASML PAS 5500/300C DUV Wafer Stepper with a resolution $<$ \SI{0.2}{\um}, serving as the oxide etch mask. A followed oxide etch in an Oxford inductively coupled plasma (ICP) dielectric etcher with \ch{O_2} and \ch{CHF_3} gases transferred the grid pattern from the photoresist to the \ch{SiO_2} layer. The photoresist residue was removed further in an \ch{O_2} asher. The other photoresist silicon etch mask was then spun and exposed by the same stepper mentioned before, shown in process (b).

After a short descum and seasoning process, we proceeded to do the first silicon etch using a Plasma-Therm DRIE deep silicon etcher. Since the etch rates decrease significantly as the plasma goes deeper\cite{Cothard2018OptimizingTE,defrance20181}, the target depth for the first silicon etch was not $D_2 - D_1$, but set to be a larger value $D'$ = \SI{137.0}{\um} instead; this initial depth was established by a few test runs. When the target depth $D'$ was reached, the wafer was cleaned with an \ch{O_2} plasma for half an hour at 3000 W to remove the photoresist and \ch{C_4 F_8} related passivation residue. Then the second silicon etch was carried out, which constructed the upper ARC layer and pushed the lower ARC layer further down to the designated depths. 

To further improve the morphology of the etched surfaces, we adopted a thermal oxidation and a clean-up process.\cite{jung2016multistep} We grew a sacrificial layer of thermal \ch{SiO_2} of about 1.1 \unit{\um} thickness using water vapors at 1200 {\textcelsius} for 100 mins in a furnace and then removed it by a hydrofluoric acid (HF) bath. The growth of the thermal \ch{SiO_2} layer and undercuts caused by previous etching processes widened the hole structure on both layers. Therefore, we shrank the dimensions on the mask designs by 1 to 2 \unit{\um} accordingly to compensate for it.

\begin{figure}
\centering
\includegraphics[width=\textwidth]{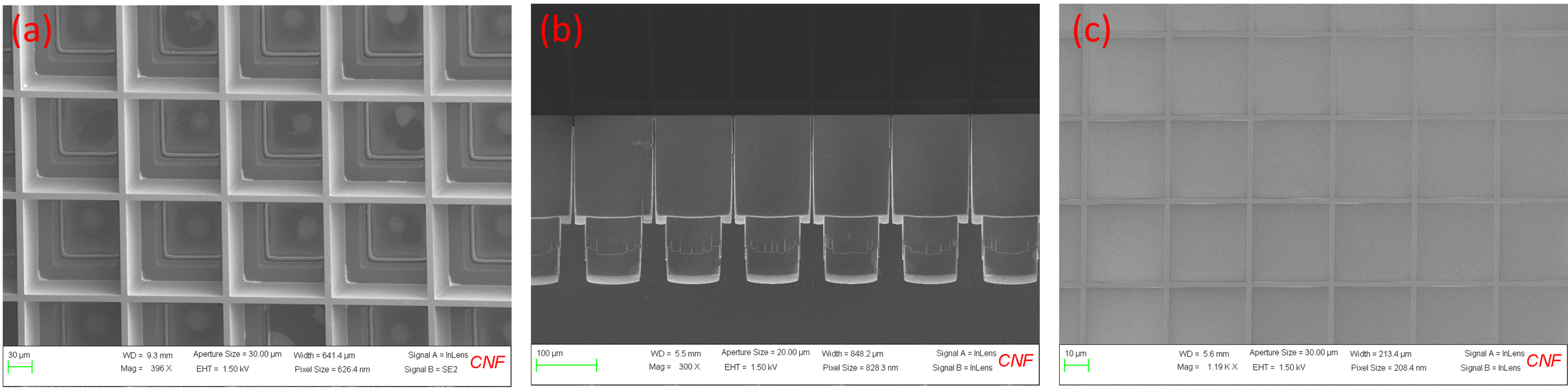}
\caption{SEM images of the double-layer ARC and metal mesh structure. (a) ARC isometric view. (b) ARC cross section view. (c) Metal mesh top view.}
\label{fig:sem}
\end{figure}

Two scanning electron microscope (SEM) images of the double-layer ARC structure are shown in Figure \ref{fig:sem}a and \ref{fig:sem}b. The whitish color in Figure \ref{fig:sem}a is caused by liquid (water, HF, and other chemicals used for cleaning up) residue on the wafer. Figure \ref{fig:sem}b also shows some irregular structures like tapered sidewalls and curved bottom surfaces.

\subsection{Metal Meshes}
After constructing the double-layer ARC structure on one side of the silicon wafer, a layer of 2 \unit{\um} photoresist was spun and baked on top of it to protect it from scratches and dust. Standard electron beam lithography, metal evaporation, and lift-off technologies were used to pattern the metal mesh structure on the other side. We coated the standard electron beam resist PMMA on the wafer and exposed it in a JEOL JBX9500FS electron beam lithography system with a maximum resolution of 6 nm. The wafer was then descumed and mounted inside a CVC SC4500 e-gun evaporation system where a 10 nm chromium adhesion layer and a 100 nm gold layer were evaporated. Afterward, we soaked the wafer into a Microposit 1165 remover bath heated to 80 {\textcelsius} to lift the gold-coated PMMA layer off. We show an SEM image for the metal mesh structure in Figure \ref{fig:sem}c.

\section{Characterization}
\label{sec:char}
\subsection{Dimension Characterization}
We cleaved each sample wafer into a 1.5 in $\times$ 1.5 in central square piece after finishing the whole fabrication process. The geometry of the ARC and metal mesh structure can be measured by examining the cleaved wafers with an SEM. 
% As mentioned in Sect.~\ref{sec:design}, the ARC structure can be characterized by the etch depths ($D_1$ and $D_2$), the square widths ($w_1$ and $w_2$) and the taper angles ($\alpha_1$ and $\alpha_2$) while the metal mesh structure can be characterized by the opening width ($w_i$). 
For one sample wafer, called the ARC mirror, we fabricated only the ARC structure without coating the metal meshes on the wafer's other side to check merely the transmittance profile of the ARC. We have two other sample pieces fabricated with both the ARC and metal mesh structure to construct one full FPI, denoted by FPI Mirror 1 and 2. Transmittance measurements for the corresponding full FPI are still ongoing. For all these three wafer pieces, we characterized physical dimensions at nine different points evenly distributed in the 1.5 in $\times$ 1.5 in square (one point at the center and the other points along four edges). Note that for the sample point at the center, we cannot measure the taper angles since the cross-section of the ARC structure is not accessible. Nevertheless, all other dimensions can be measured using a combination of an SEM and an optical profilometer. Table \ref{tab:Data} summarizes the lab measurements of the mean values and corresponding standard deviations of the parameters of all the three sample pieces discussed in the paper.

\subsection{Transmittance Measurements}

\begin{figure}
\centering
\includegraphics[width=\textwidth]{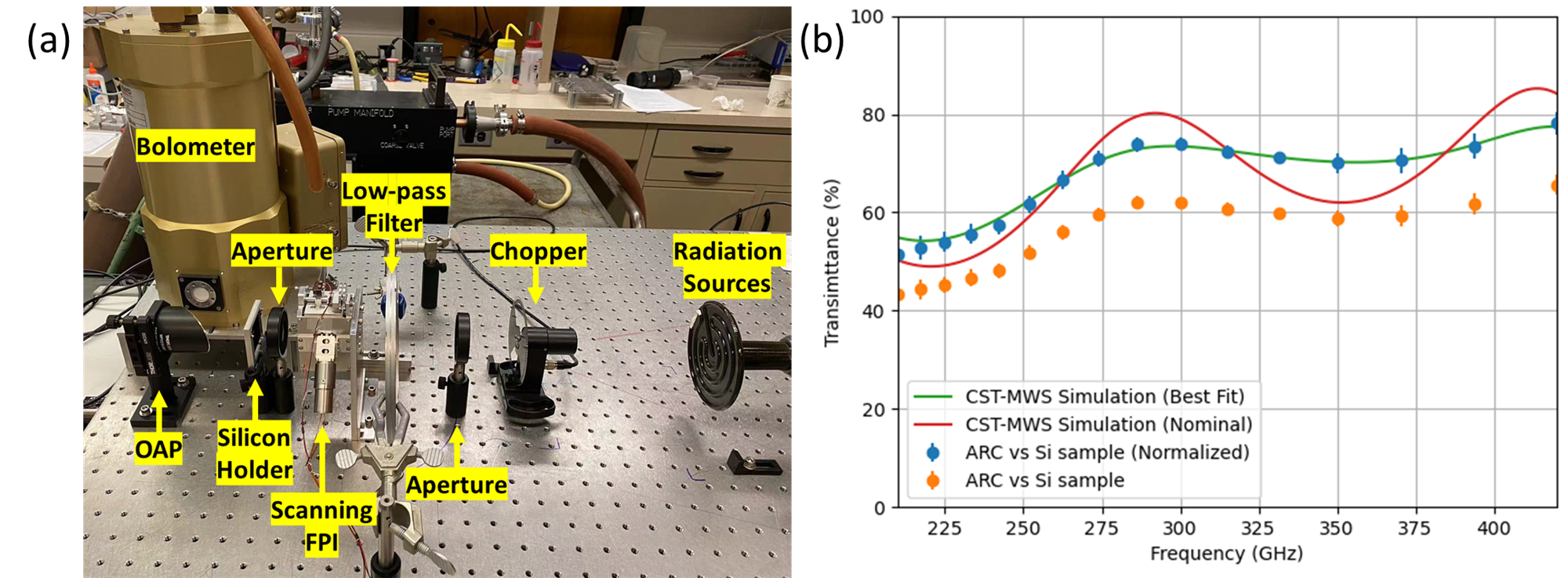}
\caption{(a) Preliminary test setup for transmittance measurements of silicon mirrors. (b) Preliminary transmittance measurements of the ARC mirror sample and corresponding CST-MWS calculations. The data points in orange are the raw measurements. They are normalized to blue dots per the 75 \% transmittance measured at 290 GHz by the laser. The red curve shows the simulation using the sample dimensions quoted in Table \ref{tab:Data}. The green curve displays the best-fit simulation using a different set of dimensions within measurement errors.}
\label{fig:char}
\end{figure}

The preliminary test setup for transmittance measurements of silicon mirrors is shown in Figure \ref{fig:char}a. The chain starts with the radiation sources of the system. Three radiation sources are available for the testing: a black body heat source (755 K), a Xenon arc lamp (5800 K) and a 290 GHz laser. The radiation is modulated by the chopper and then goes through a 1 in diameter aperture and a low-pass filter, which filters through signals with a frequency lower than 500 GHz. After that, we employ a low-order scanning FPI to sweep over the frequency band of interest. The scanning FPI comprises two free-standing metal meshes. It works in the first order and can be tuned to observe the whole band with a resolving power of about 30 via a stepper motor. The signal later passes another aperture with the same diameter and the silicon holder where the sample(s) to be measured will be mounted. An off-axis parabolic mirror (OAP) finally converges the signal to the bolometer. The output of the bolometer is further connected to a lock-in amplifier, which together with the chopper enables a synchronous detection. The Eccosorb baffles to block stray radiation are not shown in the picture. The combination of the baffles and apertures produce a nearly collimated radiation beam.

The ARC mirror was measured using the test setup above. The heat lamp was used as the radiation source. The stepper motor of the scanning FPI stepped through the whole band in sixteen equidistant steps. For each step, one sample measurement and one background measurement were taken with the sample holder in or out of the radiation beam. Then the transmittance was calculated by simply taking ratios. We also performed measurements of the ARC mirror with the 290 GHz laser. The transmittance measured at this specific frequency is 75 \%. The transmittance results of the ARC mirror and corresponding CST-MWS calculations are presented in Figure \ref{fig:char}b. The raw transmittance measurements of the ARC mirror are displayed by the orange dots. They are normalized to the blue dots by the 290 GHz laser measurements. The red curve shows the simulation using the sample dimensions quoted in Table \ref{tab:Data} while the green curve uses a different set of dimensions within measurement errors ($D_1 +\SI{1.0}{\um}, D_2 -\SI{1.0}{\um}, w_1 -\SI{0.6}{\um}, w_2 +\SI{0}{um}, \alpha_1 +\ang{0.06}, \alpha_2 +\ang{0.30}$). The good agreement between the blue dots and green curve validates the ARC model and fabrication method.

\section{Future Prospects and Additional Work in Progress}
\label{sec:future}
\subsection{Future Plan on Current Work}
We are currently updating the test setup to improve and perform transmittance measurements of both ARC vs Silicon mirrors and full SSB FPIs at room and cryogenic temperature. We will use a new set of filters in the beam path to improve the filtration against unwanted radiation signals. A dewar with a silicon holder inside will be deployed to enable cryogenic measurements. To make transmittance measurements of SSB FPIs, we plan to use two free-standing metal mesh FPIs for spectral imaging, including an order-sorting FPI and a high-order scanning FPI, to obtain sufficient resolving power. The measurements of transmittance for these samples will later be fed back to fabrication processes and simulation models.

\begin{figure}
\centering
\includegraphics[width=\textwidth]{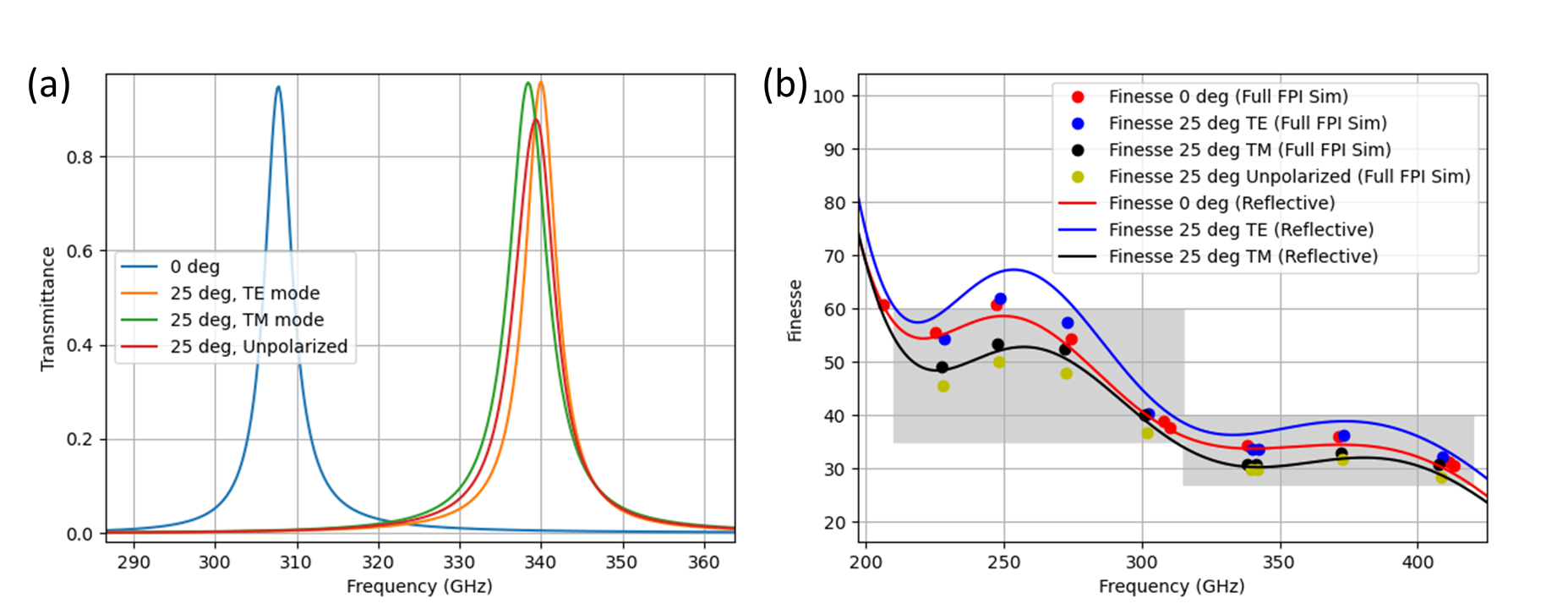}
\caption{(a) CST-MWS calculations of the transmittance fringes of the full EoR-Spec FPI with different incident angles and polarization modes. The mirror separation is set to be \SI{952}{\um}. (b) The finesse profiles and corresponding sets of finesse data points with different incident angles and polarization modes. The finesse profiles are derived from the reflectance profiles of the FPI mirror and finesse data points are calculated from the full FPI CST-MWS simulations. The grey-shaded boxes define the resolution requirement of the EoR-Spec instrument.}
\label{fig:polar}
\end{figure}

The FPI simulation model is also in development. The finesse of an FPI mirror is related to the angle and polarization mode of the incident beam. All the simulations discussed in previous sections assume normal beam incidence at the substrate surfaces. We are now taking the beam incident angle and polarization into account. Figure \ref{fig:polar}a shows the transmittance fringes of the full FPI with different incident angles and polarization modes at a fixed mirror separation. %\ang{25} is the maximum incident angle of the EoR-Spec FPI. 
Both polarization modes behave the same at normal incidence. The fringes of the FPI in TE and TM polarization modes at \ang{25} incident angle represented by the green and orange curve are blue-shifted compared to the one at normal incidence. This can be explained by the condition of maximum for an FPI, $2d \cos{\theta} = k\lambda$ where $\theta$ is the incident angle. Though not clearly seen from the plot, the FWHM and thus the finesse of these three fringes are all different. The green and orange fringes peak at different frequencies, which is caused by different phase shifts for different polarization modes. The arithmetic mean of them produces the unpolarized fringe indicated by the red curve, which will be observed by the EoR-Spec detector array since it is insensitive to polarization. The finesse and peak transmittance of the unpolarized fringe at \ang{25} is smaller than that of the fringe in either polarization mode. %That means the previous optimization for the parameters of the FPI mirror at normal incidence (Figure \ref{fig:sim2}b) may fail at larger incident angles. 
Figure \ref{fig:polar}b displays the finesse profiles and corresponding sets of finesse data points with different incident angles and polarization modes. The plot is similar to Figure \ref{fig:sim2}b, with the curves derived from the reflectance profiles of the optimized FPI mirror and data points calculated from the full FPI simulations. The curves and corresponding sets of data points roughly agree with each other with discrepancies due to simulation errors and energy losses. The olive dots representing the finesse data points for unpolarized FPI simulations at \ang{25} incident angle stay within the grey-shaded resolution targeted region, indicating that the current FPI mirror design works reasonably well for incident beams at \ang{25}. However, since the maximum incident angle of off-axis optical beams is about \ang{29},\cite{ThomasSPIE} further simulation work is needed to ensure the EoR-Spec FPI operates optimally at all incident angles. Apart from the incident angle and polarization work, we plan to model other fabrication irregularities such as rounded corners and curved bottoms on the ARCs, which can further improve the accuracy of our FPI models.

\subsection{Additional Work in Progress}
\begin{figure}
\centering
\includegraphics[scale=0.5]{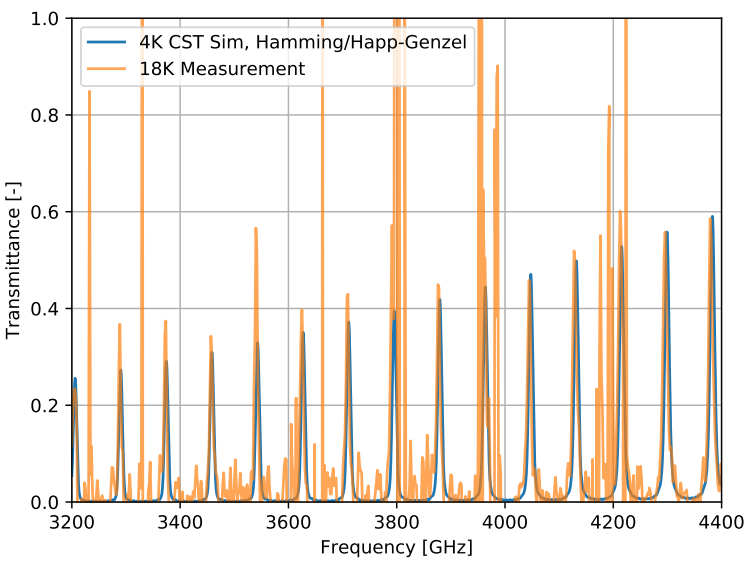}
\caption{The FTS measurements and CST-MWS simulation of a silicon-cavity fixed-FPI. The silicon substrate has a pair of inductive gold meshes on either side. The FTS data and CST-MWS calculations have been apodized with a Happ-Genzel window function.\cite{cothard2021cryogenic}}
\label{fig:fixedfpi}
\end{figure}

We also designed and fabricated silicon-cavity fixed-FPIs.\cite{Cothard2018OptimizingTE,cothard2021cryogenic} Our fixed-FPI uses a silicon substrate as the resonating cavity. The metal meshes evaporated on either side of the silicon wafer serve as metal reflectors for the FPI. It behaves as a multi-band narrow pass filter and may be used for filtering and wavelength calibration purposes. Figure \ref{fig:fixedfpi} presents the Fourier-transform spectrometer (FTS) measurements and CST-MWS simulation of a silicon-cavity fixed-FPI. The fixed-FPI consists of a \SI{525}{\um} high purity silicon substrate with inductive gold meshes patterned on either side. The detailed information about it is reported in Nicholas F. Cothard's thesis.\cite{cothard2021cryogenic} The orange part in the plot shows the 18 K FTS transmittance measurements of the fixed-FPI and the overplotted blue curve displays the corresponding CST-MWS simulation in the same frequency range at 4 K. Both parts have been apodized with a Happ-Genzel window function that we used during the FTS measurements. It shows good agreement on the widths and heights of FPI fringes between the FTS data and CST-MWS model.

Another application we are currently working on is the virtually imaged phased array (VIPA).\cite{shirasaki1999virtually,bourdarot2017nanovipa} The SSB VIPA is a high resolution, wavelength dispersion device composed of metal reflecting surfaces on either side of a thick silicon substrate piece. The front surface of the substrate is coated with a nearly 100\% reflective film (for example, gold) except for an entrance window. The rear surface is patterned with a partially reflective metal mesh. The light source is line-focused into the entrance window by a cylindrical lens at near-normal incidence and bounces around within the silicon cavity resembling an FPI etalon. The output light from the rear surface presents a wavelength dispersion response due to interferences of these multiple reflecting beams. We have fabricated a prototype SSB VIPA piece with a 10 mm $\times$ 30 mm $\times$ 50 mm thick high purity silicon block using our metal mesh fabrication techniques and plan to characterize its spectral performance in the near future.

\section{Conclusions}
We have presented the models for ARCs, metal meshes, and an optimized design for the EoR-Spec FPI. We have also demonstrated the fabrication of the silicon mirrors for the EoR-Spec FPI. An ARC mirror was characterized by our preliminary test setup and shows good agreement with the electromagnetic simulation. Further improvements on the test setup and simulation models are ongoing. We are actively applying our SSB fabrication techniques to other optical devices including fixed-FPIs and VIPAs.

\acknowledgments % equivalent to \section*{ACKNOWLEDGMENTS}
This work is supported by NSF Grant AST-2009767, and NASA grant NNX16AC72G. The CCAT-prime project, FYST and Prime-Cam instrument have been supported by generous contributions from the Fred M. Young, Jr. Charitable Trust, Cornell University, and the Canada Foundation for Innovation and the Provinces of Ontario, Alberta, and British Columbia. The construction of the FYST telescope was supported by the Gro{\ss}ger{\"a}te-Programm of the German Science Foundation (Deutsche Forschungsgemeinschaft, DFG) under grant INST 216/733-1 FUGG, as well as funding from Universit{\"a}t zu K{\"o}ln, Universit{\"a}t Bonn and the Max Planck Institut f{\"u}r Astrophysik, Garching. The construction of EoR-Spec is supported by NSF grant AST-2009767. This work was performed in part at the Cornell NanoScale Facility, a member of the National Nanotechnology Coordinated Infrastructure (NNCI), which is supported by the National Science Foundation (Grant NNCI-2025233). SKC acknowledges support from NSF award AST-2001866. ZBH acknowledges support from a NASA Space Technology Graduate Research Opportunities Award.

% References
\bibliography{report} % bibliography data in report.bib
\bibliographystyle{spiebib} % makes bibtex use spiebib.bst

\end{document}